\input{aipcheck}

\documentclass[
    ,final            
  ]
  {aipproc}

\layoutstyle{6x9}

\usepackage{epsfig,float}

\newcommand{\be}{\begin{equation}}
\newcommand{\ee}{\end{equation}}
\newcommand{\ba}{\begin{eqnarray}}
\newcommand{\ea}{\end{eqnarray}}

\begin{document}

\title{Dynamically generated hadron resonances}


\author{E. Oset, E. J. Garzon, Ju Jun Xie, P. Gonzalez}{
   address={Departamento de Fisica Teorica and IFIC, Centro Mixto Universidad de Valencia-CSIC,
Institutos de Investigacion de Paterna, Aptd. 22085, 46071 Valencia, Spain.}
}

\author{A. Ramos}{
  address={Departament d'Estructura i Constituents de la Materia, Universitat de
Barcelona}
}

\author{A. Mart\'inez Torres}{
  address={Yukawa Institute for Theoretical Physics, Kyoto University, Kyoto 606-8502, Japan}
}

\begin{abstract}  
As an example of dynamically generated resonances we mention the interaction of vector mesons with baryons within the local hidden gauge formalism which gives rise to a large amount of such states, many of which can be
associated to known resonances, while others represent predictions for new
resonances. The width of these states coming from decay into pseudoscalar baryon is also addressed. We also mention recent states coming from $\Delta \rho \pi$ interaction obtained with Faddeev equations.
\end{abstract}

\maketitle


\section{Introduction}

The combination of effective Lagrangians to account for hadron interaction with a coupled channel approach, implementing exactly unitarity in coupled channels, has turned out to be a very efficient tool to 
 face many problems in Hadron Physics. Using this so called coupled channel unitary approach, usually referred to as chiral unitary approach, since the Lagrangians used account for chiral symmetry, the 
interaction of the octet of
pseudoscalar mesons with the octet of stable baryons has been studied and 
leads to $J^P=1/2^-$
resonances which fit quite well the spectrum of the known low lying resonances
with these quantum numbers 
\citep{Kaiser:1995cy,angels,ollerulf,carmenjuan,hyodo}. 
 New resonances are sometimes predicted, the most notable
being the $\Lambda(1405)$, where all the
chiral approaches find two poles close by 
\citep{Jido:2003cb,Borasoy:2005ie,Oller:2005ig,Oller:2006jw,Borasoy:2006sr,Hyodo:2008xr,Roca:2008kr}, rather than one, for which 
experimental support is presented in \citep{magas,sekihara}.  Another step forward in this
 direction has been the interpretation
of low lying $J^P=1/2^+$ states as molecular systems of two pseudoscalar mesons and one baryon
\citep{alberto,alberto2,kanchan,Jido:2008zz,KanadaEn'yo:2008wm}. 

Much work has been done using pseudoscalar
mesons as building blocks, but more recently, vectors instead of
pseudoscalars are also being considered. In the baryon sector the
interaction of the $\rho \Delta$ interaction has been recently addressed in
\citep{vijande}, where three degenerate $N^*$ states and three degenerate
$\Delta$ states around 1900 MeV, with $J^P=1/2^-, 3/2^-, 5/2^-$, are found. The extrapolation to SU(3) with the interaction of the vectors of the nonet with
the baryons of the decuplet has been done in \citep{sourav}. The
underlying theory for this study is the hidden gauge formalism
\citep{hidden1,hidden2,hidden4}, which deals with the interaction of vector mesons and
pseudoscalars, respecting chiral dynamics, providing the interaction of
pseudoscalars among themselves, with vector mesons, and vector mesons among
themselves. It also offers a perspective on the chiral Lagrangians as limiting
cases at low energies of vector exchange diagrams occurring in the theory.

 In the meson sector, the interaction of $\rho \rho$ within this formalism has
been addressed in \citep{raquel}, where it has been shown to lead  to the
dynamical generation of the $f_2(1270)$ and $f_0(1370)$  meson resonances. The extrapolation to SU(3) of the work 
of \citep{raquel} has been done in \citep{gengvec}, where many resonances are
obtained, some of which can be associated to known meson states, while there are
predictions for new ones.

  In this talk we present the results of the interaction of the nonet
   of vector mesons  with the
   octet of baryons \citep{angelsvec}, which have been done 
 using the unitary approach in coupled
channels. The scattering amplitudes lead to poles in the
complex plane which can be associated to some well known resonances. Under the
approximation of neglecting the three momentum of the particles versus their
mass, we obtain degenerate states of $J^P=1/2^-,3/2^-$ for the case of the
interaction of vectors with the octet of baryons. This degeneracy 
seems to be followed qualitatively by the experimental spectrum, although in
some cases the spin partners have not been identified. Improvements in the theory follow from the consideration of the decay of these states into a pseudoscalar meson and a baryon, and some results are presented here.
 Finally we shall also report on some states coming from the three body system 
 $\Delta \rho \pi$ which can shed some light on the status of some $\Delta$ states of $J^P=5/2^+$ in the vicinity of 2000 MeV.

\section{Formalism for $VV$ interaction}

We follow the formalism of the hidden gauge interaction for vector mesons of 
\citep{hidden1,hidden2} (see also \citep{hidekoroca} for a practical set of Feynman rules). 
The Lagrangian involving the interaction of 
vector mesons amongst themselves is given by
\begin{equation}
{\cal L}_{III}=-\frac{1}{4}\langle V_{\mu \nu}V^{\mu\nu}\rangle \ ,
\label{lVV}
\end{equation}
where the symbol $\langle \rangle$ stands for the trace in the SU(3) space 
and $V_{\mu\nu}$ is given by 
\begin{equation}
V_{\mu\nu}=\partial_{\mu} V_\nu -\partial_\nu V_\mu -ig[V_\mu,V_\nu]\ ,
\label{Vmunu}
\end{equation}
with  $g$ given by $g=\frac{M_V}{2f}$
where $f=93\,MeV$ is the pion decay constant. The magnitude $V_\mu$ is the SU(3) 
matrix of the vectors of the nonet of the $\rho$
\begin{equation}
V_\mu=\left(
\begin{array}{ccc}
\frac{\rho^0}{\sqrt{2}}+\frac{\omega}{\sqrt{2}}&\rho^+& K^{*+}\\
\rho^-& -\frac{\rho^0}{\sqrt{2}}+\frac{\omega}{\sqrt{2}}&K^{*0}\\
K^{*-}& \bar{K}^{*0}&\phi\\
\end{array}
\right)_\mu \ .
\label{Vmu}
\end{equation}

The interaction of ${\cal L}_{III}$ gives rise to a contact term coming from 
$[V_\mu,V_\nu][V_\mu,V_\nu]$
\begin{equation}
{\cal L}^{(c)}_{III}=\frac{g^2}{2}\langle V_\mu V_\nu V^\mu V^\nu-V_\nu V_\mu
V^\mu V^\nu\rangle\ ,
\label{lcont}
\end{equation}
 and on the other hand it gives rise to a three 
vector vertex from 
\begin{equation}
{\cal L}^{(3V)}_{III}=ig\langle (\partial_\mu V_\nu -\partial_\nu V_\mu) V^\mu V^\nu\rangle
\label{l3V}=ig\langle (V^\mu\partial_\nu V_\mu -\partial_\nu V_\mu
V^\mu) V^\nu\rangle
\label{l3Vsimp}\ ,
\end{equation}

In this latter case one finds an analogy with the coupling of vectors to
 pseudoscalars given in the same theory by 
 
\be
{\cal L}_{VPP}= -ig ~tr\left([
P,\partial_{\mu}P]V^{\mu}\right),
\label{lagrVpp}
\ee
where $P$ is the SU(3) matrix of the pseudoscalar fields. 

In a similar way, we have the Lagrangian for the coupling of vector mesons to
the baryon octet given by
\citep{Klingl:1997kf,Palomar:2002hk}

\be
{\cal L}_{BBV} =
\frac{g}{2}\left(tr(\bar{B}\gamma_{\mu}[V^{\mu},B])+tr(\bar{B}\gamma_{\mu}B)tr(V^{\mu})
\right),
\label{lagr82}
\ee
where $B$ is now the SU(3) matrix of the baryon octet \citep{Eck95,Be95}. Similarly,
one has also a lagrangian for the coupling of the vector mesons to the baryons
of the decuplet, which can be found in \citep{manohar}.

\begin{figure}[tb]
\epsfig{file=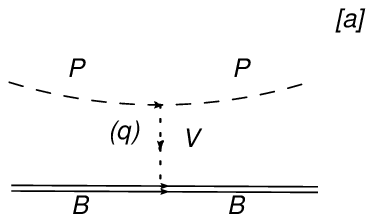, width=7cm} \epsfig{file=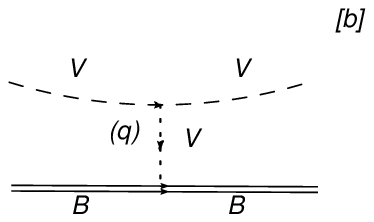, width=7cm}
\caption{Diagrams obtained in the effective chiral Lagrangians for interaction
of pseudoscalar [a] or vector [b] mesons with the octet or decuplet of baryons.}%
\label{f1}%
\end{figure}

With these ingredients we can construct the Feynman diagrams that lead to the $PB
\to PB$ and $VB \to VB$ interaction, by exchanging a vector meson between the
pseudoscalar or the vector meson and the baryon, as depicted in Fig.\ref{f1} .

 As shown in \citep{angelsvec}, in the limit of small three momenta of the vector mesons, which we consider, the vertices of Eq. (\ref{l3Vsimp}) and Eq. (\ref{lagrVpp}) give rise to the same expression.  This makes the work technically easy allowing the use of many previous results.

   A small amendment is in order in the case of vector mesons, which
   is due to the mixing of $\omega_8$ and the singlet of SU(3), $\omega_1$, to give the
   physical states of the $\omega$ and the $\phi$.
    In this case, all one must do is to take the
   matrix elements known for the $PB$ interaction and, wherever $P$ is the
   $\eta_8$, multiply the amplitude by the factor $1/\sqrt 3$ to get the
   corresponding $\omega $ contribution and by $-\sqrt {2/3}$ to get the
   corresponding $\phi$ contribution.  Upon the approximation consistent with
   the neglect of the three momentum versus the mass of the particles (in this
   case the baryon), we can just take the $\gamma^0$ component of 
   eq. (\ref{lagr82})  and
   then the transition potential corresponding to the diagram of Fig. 1(b ) is
   given by
   
   \begin{equation}
V_{i j}= - C_{i j} \, \frac{1}{4 f^2} \, (k^0 + k'^0)~ \vec{\epsilon}\vec{\epsilon
} ',
\label{kernel}
\end{equation}
 where $k^0, k'^0$ are the energies of the incoming and outgoing vector mesons. 
   The same occurs in the case of the decuplet.  
    
    The $C_{ij}$ coefficients of Eq. (\ref{kernel}) can be obtained directly from 
    \citep{angels,bennhold,inoue}
    with the simple rules given above for the $\omega$ and the $\phi$, and
    substituting $\pi$ by $\rho$ and $K$ by $K^*$ in the matrix elements. The
    coefficients are obtained both in the physical basis of states or in the
    isospin basis. Here we will show results in isospin
    basis. 

    The next step to construct the scattering matrix is done by solving the
    coupled channels Bethe Salpeter equation in the on shell factorization approach of 
    \citep{angels,ollerulf}
   \begin{equation}
T = [1 - V \, G]^{-1}\, V,
\label{eq:Bethe}
\end{equation} 
with $G$ the loop function of a vector meson and a baryon which we calculate in
dimensional regularization using the formula of \citep{ollerulf} and similar
values for the subtraction constants. The $G$ function is convoluted with the 
spectral function for the vector mesons to take into account their width.

 The iteration of diagrams implicit in the Bethe Salpeter equation in the case
 of the vector mesons propagates the $\vec{\epsilon}\vec{\epsilon }'$ term 
 of the interaction, thus,
the factor $\vec{\epsilon}\vec{\epsilon }'$ appearing in the potential $V$,
factorizes also in the $T$ matrix for the external vector mesons. This has as a consequence that the interaction is spin independent and we find degenerate states in $J^P=1/2$ and $J^P=3/2$.

\section{Results} 

 The resonances obtained are summarized in Table  \ref{tab:octet} .
 As one can see in Table \ref{tab:octet} there are states which one can easily
associate to known resonances. There are ambiguities in other cases. One can also
see that in several cases the degeneracy in spin that the theory predicts is
clearly visible in the experimental data, meaning that there are 
several states with about 50 MeV 
or less mass difference
between them.  In some cases, the theory predicts quantum numbers for 
resonances which have no spin and parity associated. It would be interesting to
pursue the experimental research to test the theoretical predictions.

\begin{table}[ht]
      \renewcommand{\arraystretch}{1.5}
     \setlength{\tabcolsep}{0.2cm}
\begin{tabular}{c|c|cc|ccccc}\hline\hline
$S,\,I$&\multicolumn{3}{c|}{Theory} & \multicolumn{5}{c}{PDG data}\\
\hline
    \vspace*{-0.3cm}
    & pole position    & \multicolumn{2}{c|}{real axis} &  &  & &  &  \\
    & {\small (convolution)}    &\multicolumn{2}{c|}{{\small
    (convolution)}} & \\
    &   & mass & width &name & $J^P$ & status & mass & width \\
    \hline
$0,1/2$ & --- & 1696  & 92  & $N(1650)$ & $1/2^-$ & $\star\star\star\star$ & 1645-1670
& 145-185\\
  &      &       &     & $N(1700)$ & $3/2^-$ & $\star\star\star$ &
	1650-1750 & 50-150\\
       & $1977 + {\rm i} 53$  & 1972  & 64  & $N(2080)$ & $3/2^-$ & $\star\star$ & $\approx 2080$
& 180-450 \\	
   &     &       &     & $N(2090)$ & $1/2^-$ & $\star$ &
 $\approx 2090$ & 100-400 \\
 \hline
$-1,0$ & $1784 + {\rm i} 4$ & 1783  & 9  & $\Lambda(1690)$ & $3/2^-$ & $\star\star\star\star$ &
1685-1695 & 50-70 \\
  &       &       &    & $\Lambda(1800)$ & $1/2^-$ & $\star\star\star$ &
1720-1850 & 200-400 \\
       & $1907 + {\rm i} 70$ & 1900  & 54  & $\Lambda(2000)$ & $?^?$ & $\star$ & $\approx 2000$
& 73-240\\
       & $2158 + {\rm i} 13$ & 2158  & 23  &  &  &  & & \\
       \hline
$-1,1$ & $ --- $ & 1830  & 42  & $\Sigma(1750)$ & $1/2^-$ & $\star\star\star$ &
1730-1800 & 60-160 \\
  & $ --- $   & 1987  & 240  & $\Sigma(1940)$ & $3/2^-$ & $\star\star\star$ & 1900-1950
& 150-300\\
   &     &       &   & $\Sigma(2000)$ & $1/2^-$ & $\star$ &
$\approx 2000$ & 100-450 \\\hline
$-2,1/2$ & $2039 + {\rm i} 67$ & 2039  & 64  & $\Xi(1950)$ & $?^?$ & $\star\star\star$ &
$1950\pm15$ & $60\pm 20$ \\
         & $2083 + {\rm i} 31 $ &  2077     & 29  &  $\Xi(2120)$ & $?^?$ & $\star$ &
$\approx 2120$ & 25  \\
 \hline\hline
    \end{tabular}
\caption{The properties of the 9 dynamically generated resonances stemming from the vector-baryon octet interaction and their possible PDG
counterparts.}
\label{tab:octet}
\end{table}

  The predictions made here for resonances not observed should be a stimulus for
further search of such states. In this
sense it is worth noting the experimental program at Jefferson Lab 
\citep{Price:2004xm} to investigate the $\Xi$ resonances. We are
confident that the predictions  shown here stand on solid grounds and anticipate much
progress in the area of baryon spectroscopy and on the understanding of the
nature of the baryonic resonances. 

\section{Incorporating the pseudoscalar meson-baryon channels}

\begin{center}
\begin{figure}[h!]
\begin{tabular}{ccc}
\includegraphics[scale=0.5]{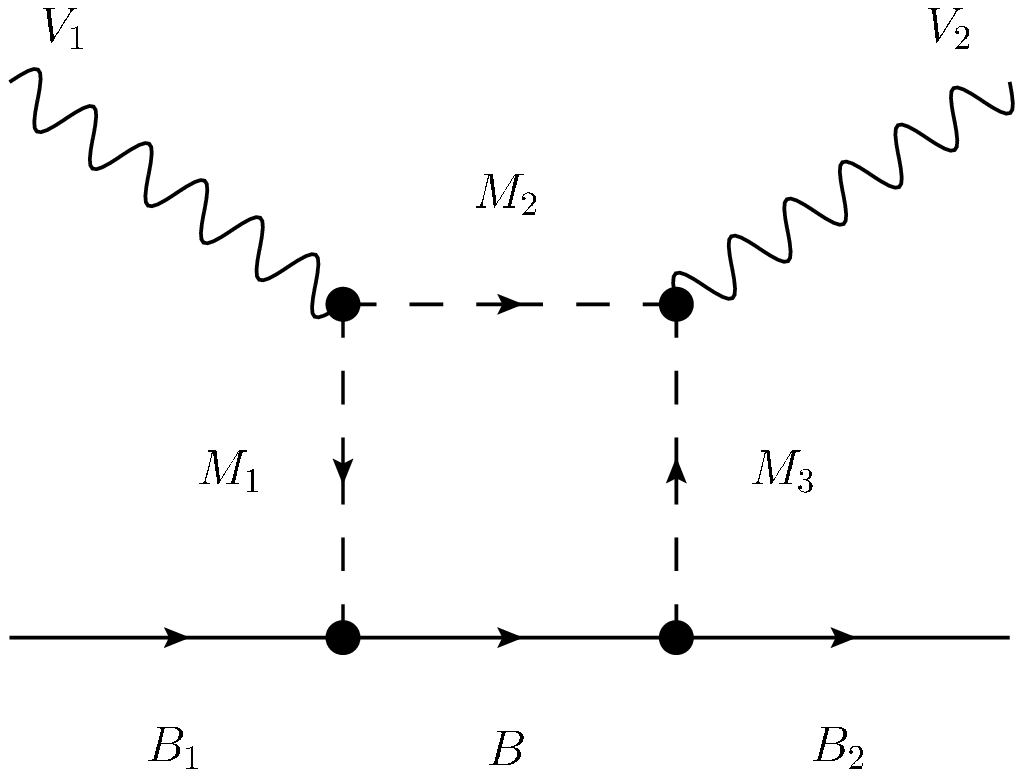} &~~~~& \includegraphics[scale=0.5]{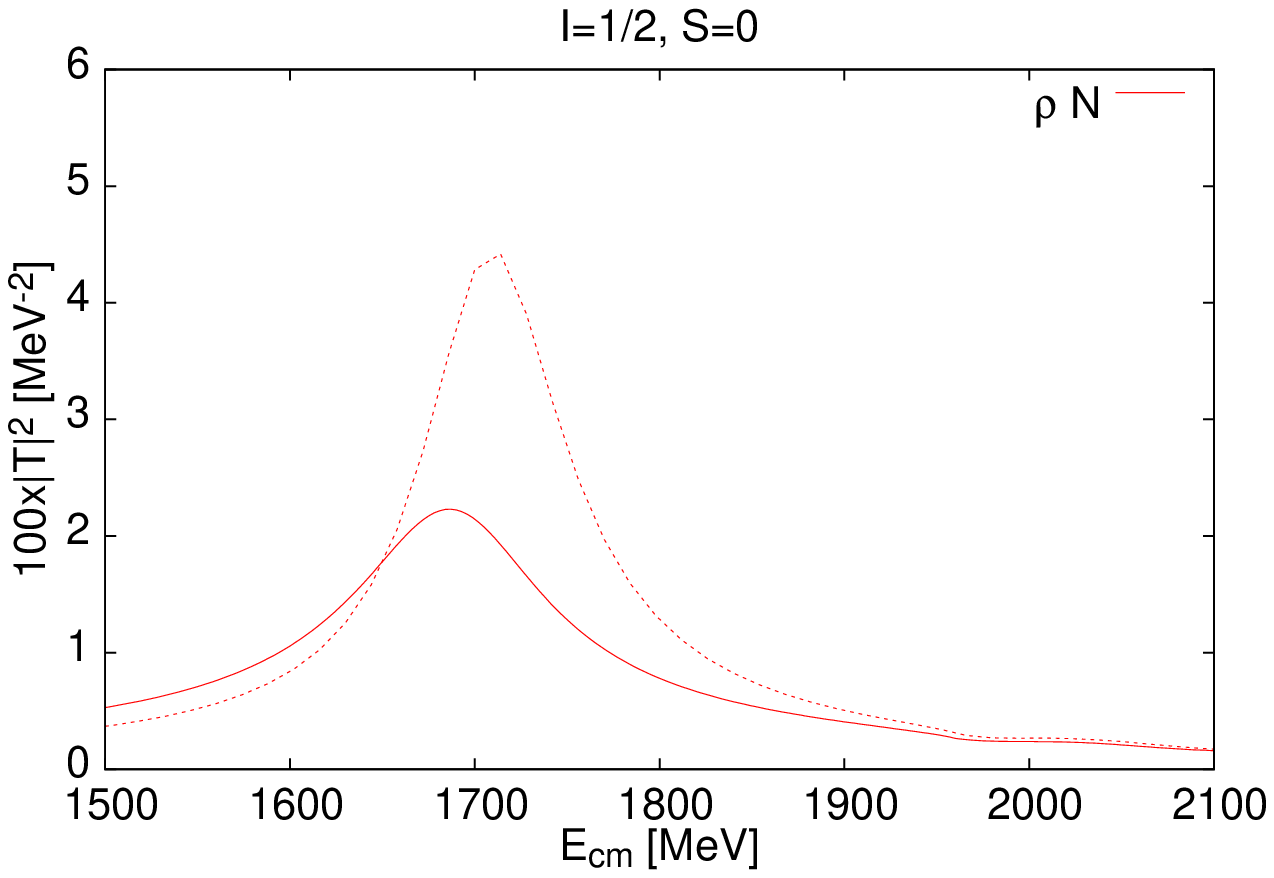} \\
a &~~~~& b
\end{tabular}
\caption{a) Diagrams for $VB \rightarrow VB$ interaction incorporating the intermediate pseudoscalar-baryon states. b) Result for the $\rho N$ channel: dashed) tree level of vector-baryon interaction, solid) including intermediate pseudoscalar-baryon states.}
\label{box}
\end{figure}
\end{center}

Improvements in the states tabulated in Table \ref{tab:octet} have been done by incorporating intermediate states of a pseudoscalar meson and a baryon \citep{garzon}. This is done by including the diagrams of Fig.~(\ref{box}).

In the intermediate B states of Fig. 2(a) we include baryons of the octet and the decuplet. The results of the calculations are a small shift and a broadening of the resonances obtained with the base of vector-baryon alone. In Fig.~2(b) we show some of the results obtained. In particular it is interesting to see that a $N^*$ resonance appearing around 1700 MeV in the $\rho N$ channel moves to lower energies and becomes wider, to the point that it could be associated to the $N^*(1650)$ $1/2^-$, which does not appear in ordinary chiral unitary approaches where only pseudoscalar meson-baryon states are considered as building blocks.

\section{The $\Delta \rho \pi$ system and $\Delta$  $J^P=5/2^+$ states around 2000 MeV}
In Refs.~\citep{vijande,sourav} it show that the $\Delta \rho$ interaction
gave rise to $N^*$ and $\Delta$ states with degenerate spin-parity
$1/2^-,3/2^-,5/2^-$. In a recent work~\citep{Xie:2011ab}, one extra
$\pi$ was introduced in the system, and via the Fixed Center
Approximation to the Faddeev Equations, the new system was studied
and new states were found.

\begin{center}%
\begin{table}[ptbh]
\caption{Assignment of $I=3/2$ predicted states to $J^{P}=1/2^{+}%
,3/2^{+},5/2^{+}$ resonances. Estimated PDG masses for these
resonances as well as their extracted values from references
\citep{Man92} and \citep{Vra00} (in brackets) are shown for
comparison. N. C. stands for a non cataloged resonance in the PDG
review}
\begin{tabular}
[c]{c|ccccc} \hline
Predicted & \multicolumn{5}{c}{PDG data}\\
&  &  &  &  & \\
Mass (MeV) & Name & $J^{P}$ & Estimated Mass (MeV) & Extracted Mass (MeV) & Status\\
1770 & $\Delta(1740)$ & $5/2^{+}$ &  & $1752\pm32$ & N.C.\\
&  &  &  & $(1724\pm61)$ & \\
& $\Delta(1600)$ & $3/2^{+}$ & $1550-1700$ & $1706\pm10$ & *** \\
&  &  &  & $(1687\pm44)$ & \\
& $\Delta(1750)$ & $1/2^{+}$ & $\approx1750$ & $1744\pm36$ & * \\
&  &  &  & $(1721\pm61)$ & \\\hline
$1875$ & $\Delta(1905)$ & $5/2^{+}$ & $1865-1915$ & $1881\pm18$ & ****\\
&  &  &  & $(1873\pm77)$ & \\
& $\Delta(1920)$ & $3/2^{+}$ & $1900-1970$ & $2014\pm16$ & ***\\
&  &  &  & $(1889\pm100)$ & \\
& $\Delta(1910)$ & $1/2^{+}$ &
$1870-1920$ & $1882\pm10$ & ****\\
&  &  &  & $(1995\pm12)$ &
\\\hline
\end{tabular}
\end{table}
\end{center}

We show in Table 2 the two $\Delta^*$ states obtained and
there is also hint of another $\Delta^*$ state around $2200$ MeV.
Experimentally, only two resonances $\Delta_{5/2^+}(1905)(****)$ and
$\Delta_{5/2^+}(2000)(**)$, are cataloged in the Particle Data Book
Review~\citep{pdg2010}. However, a careful look at
$\Delta_{5/2^+}(2000)(**)$, shows that its nominal mass is in fact
estimated from the mass $(1724\pm61)$, $(1752\pm32)$ and
$(2200\pm125)$ respectively, extracted from three independent
analyses of different character~\citep{Man92,Vra00,Cut80}. Moreover a
recent new data analysis~\citep{Suz10} has reported a
$\Delta_{5/2^+}$ with a pole position at $1738$ MeV.

Our results give quantitative theoretical support to the existence
of two distinctive resonances, $\Delta_{5/2^+}(\sim 1740)$ and
$\Delta_{5/2}^+(\sim 2200)$. We propose that these two resonances
should be cataloged instead of $\Delta_{5/2^+}(2000)$. This proposal
gets further support from the possible assignment of the other
calculated baryon states in the $I=1/2,3/2$ and $J^P=1/2^+,3/2^+$
sectors to known baryonic resonances. In particular the poorly
established $\Delta_{1/2^+}(1750)(*)$ may be naturally interpreted
as a $\pi N_{1/2^-}(1650)$ bound state.

\section*{Acknowledgments}

This work is partly supported by DGICYT contract number
FIS2006-03438.
This research is  part of the EU Integrated Infrastructure Initiative Hadron Physics Project
under  contract number RII3-CT-2004-506078.

\end{document}